\documentclass[conference]{IEEEtran}
\usepackage[pdftex]{graphicx}
   \graphicspath{{../pdf/}{../jpeg/}}
	\DeclareGraphicsExtensions{.pdf,.jpeg,.png}

	\usepackage[cmex10]{amsmath}
	\usepackage{mathabx}
	\usepackage{algorithmic}
	\usepackage{array}
	\usepackage{mdwmath}
	\usepackage{mdwtab}
	\usepackage{eqparbox}
    \newcolumntype{P}[1]{>{\centering\arraybackslash}p{#1}}
    \newcolumntype{M}[1]{>{\centering\arraybackslash}m{#1}}
    \usepackage{rotating}
 
    \usepackage{tabularx}
    \usepackage{caption}
    \usepackage{mwe}
    \usepackage{graphicx}
    \usepackage{sidecap}
    \usepackage{derivative}
    \usepackage{amsmath}
    \usepackage{cite}
  
    \DeclareMathOperator{\taninv}{tan^{-1}}
    
\begin{document}

\title{\LARGE  Optimisation of design parameters to improve performance of a planar electromagnetic actuator}

\author{ K. S. Vikrant, D. Dadkhah, and S. O. Reza Moheimani, Fellow, IEEE }

\maketitle

\begin{abstract}
Planar electromagnetic actuators based on the principle of linear motors are widely employed for micro and nano positioning applications. These actuators usually employ a planar magnetic platform driven by a co-planar electromagnetic coil. While these actuators offer a large motion range and high positioning resolution, their actuation bandwidth is limited due to relatively small electromagnetic stiffness.  We report optimization of the design parameters of the electromagnetic coil and the magnetic assembly to maximize the electromagnetic force and stiffness.  Firstly, we derive closed-form expressions for the electromagnetic forces and stiffness, which enable us to express these quantities in terms of the design parameters of the actuator. Secondly, based on these derived expressions, we estimate the optimum values of the design parameters to maximize force and stiffness. Notably, for the optimum design parameters, the force and stiffness per unit volume can be increased by two and three orders of magnitude, respectively by reducing the pitch of the electromagnetic coil by a factor of 10.  Lastly, we develop an electromagnetic actuator and evaluate its performance using a Microelectromechanical system (MEMS) based force sensor. By operating the force sensor in a feedback loop, we precisely measure the generated electromagnetic forces for different design parameters of the actuator. The experimental results obtained align closely with the analytical values, with an error of less than 15\%.

\end{abstract}

\IEEEoverridecommandlockouts


\IEEEpeerreviewmaketitle

\section{Introduction}
Electromagnetic actuators have been widely utilized in motion control applications for almost two centuries due to their precision motion characteristics, high load-driving capability, large bandwidth, simpler control, and high efficiency\cite{8772178,boldea2004linear,saidur2010review,6980715}. These actuators can generate both torque and force, enabling rotational and linear motion, respectively. Their numerous advantages make electromagnetic actuators a viable candidate for micro and nanotechnology applications. However, conventional designs are unsuitable for developing micro-scale electromagnetic actuators due to two specific challenges.

The first challenge emerges due to the friction within the sliding parts such as lead screws, bearing and guides used in the conventional electromagnetic actuator. As the dimensions are scaled down, the surface-to-volume ratio increases, resulting in higher frictional forces and reduced electromagnetic forces. Consequently, the efficiency and positioning accuracy of micro-scale electromagnetic actuators is compromised \cite{smith2006sensor,4141033,bell2005mems}. The second challenge arises from the complexity of designing miniature coils and magnetic assemblies using available fabrication techniques\cite{niarchos2003magnetic,800629}.   

To address the first challenge, researchers have proposed two different designs of electromagnetic actuators without mechanical transmission elements. The first design suspends the magnetic assembly with flexure-guided beams, providing one or more degrees of freedom\cite{ahn2015design,ehle2013moving}. However, this design limits the motion range. The second design utilizes an untethered magnetic assembly, either immersed in a liquid or levitated in the air, to eliminate friction during motion. The suspended, immersed, or levitated magnetic assembly is actuated using current-carrying coils, enabling smooth large range motion\cite{pawashe2009modeling,uvet2018micro,zhou2022magnetic}.

To address the second challenge, simpler designs of electromagnetic coils and permanent magnets have been explored. Initially, 3D designs based on small cylindrical coils and spherical magnetic beads were proposed, allowing the development of mesoscale actuators for miniature motors, robots, coordinate measurement machines, and force sensors\cite{kohlmeier2004investigation,mrinalini2017design,narendra2022electromagnetically,jing2018microforce}. However, fabricating even these basic 3D designs at the micro-scale is challenging using conventional techniques. To overcome this limitation, researchers proposed novel 2D designs of electromagnetic coils and permanent magnet assemblies \cite{zhi2014planar,wang2021micro,8808804}. These 2D designs can be fabricated using silicon-based micro-machining techniques commonly employed for Micro-Electro-Mechanical Systems (MEMS) development.

Two prevalent designs of planar electromagnetic coils are employed in developing miniature electromagnetic actuators. The first design consists of a circular or spiral current-carrying track, primarily used for actuating flexure-guided permanent magnets and membranes. These actuators, with their large force and small motion range, are suitable for devices like micro-pumps, micro-valves, and micro-inertial switches \cite{waldschik2010micro,luharuka2008simulated,wang2022bi}. Additionally, circular or spiral tracks are employed to untetherly actuate micro-scale permanent magnet beads inside a liquid medium\cite{punyabrahma2022micro}. However, in such cases, the coils are moved by an external positioner while the magnetic bead remains electromagnetically trapped at the centre of the coil.

To achieve untethered motion without relying on an external positioner, a planar electromagnetic actuator based on the principle of a linear motor has been introduced. This actuator utilizes a planar coil and a thin magnetic stage. The planar coil consists of two independent parallel tracks arranged in a serpentine pattern. By varying the currents suitably, the position of the electromagnetically trapped magnetic stage can be precisely adjusted. Moreover, the planar coil's configuration enables the integration of multiple thin permanent magnets in a chequerboard pattern, resulting in a co-planar magnetic stage with scaled electromagnetic stiffness and force \cite{9143422,khan2011long,khan2017micro,vikrant2022diamagnetically,hsu2017automated,vikrant2023novel}. 
 
While this reported planar electromagnetic actuator exhibits several advantages, including a large motion range, small footprint, and multi-degree of freedom actuation capability, it possesses limited electromagnetic force and stiffness for given actuation currents and the magnetic stage's volume. These limitations restrict the bandwidth of levitation-based positioners to a few tens of hertz, even with high actuation currents \cite{vikrant2022diamagnetically,vikrant2023novel}. Additionally, the small electromagnetic force limits the dynamic range of flexure-guided magnetic assemblies. Overcoming these limitations can lead to high-bandwidth levitation-based nanopositioners, high dynamic range force transducers, and broaden the applicability of electromagnetic actuators in other MEMS devices where large forces and stiffness are required.
 
 In this paper, we present the optimization of the design parameters of a planar electromagnetic actuator to increase the electromagnetic force and trapping stiffness for a given actuation current and volume of the magnetic stage. We develop an analytical model of the actuator, utilize it to determine the maximum force and stiffness for optimized design parameters, and then fabricate the planar electromagnetic actuator. Integration with a MEMS-based force sensor enables us to measure the generated electromagnetic force and stiffness using a feedback loop. We compare these measured forces with analytically obtained values for different design parameters and achieve a maximum error of less than 15\%.

The structure of the remainder of this paper is as follows: Section II presents the electromagnetic modeling approach, Section III describes the development and evaluation of the planar actuator, and finally, Section IV concludes the paper.

\section{Electromagnetic modeling}

Section II-A describes the design of the actuator and the analytical model developed for calculating electromagnetic force and stiffness.  Section II-B discusses the optimisation of design parameters to increase the force and stiffness.   

\subsection{Analytical Model}

\begin{figure}
    \begin{centering}
        \includegraphics[width=0.4\paperwidth]{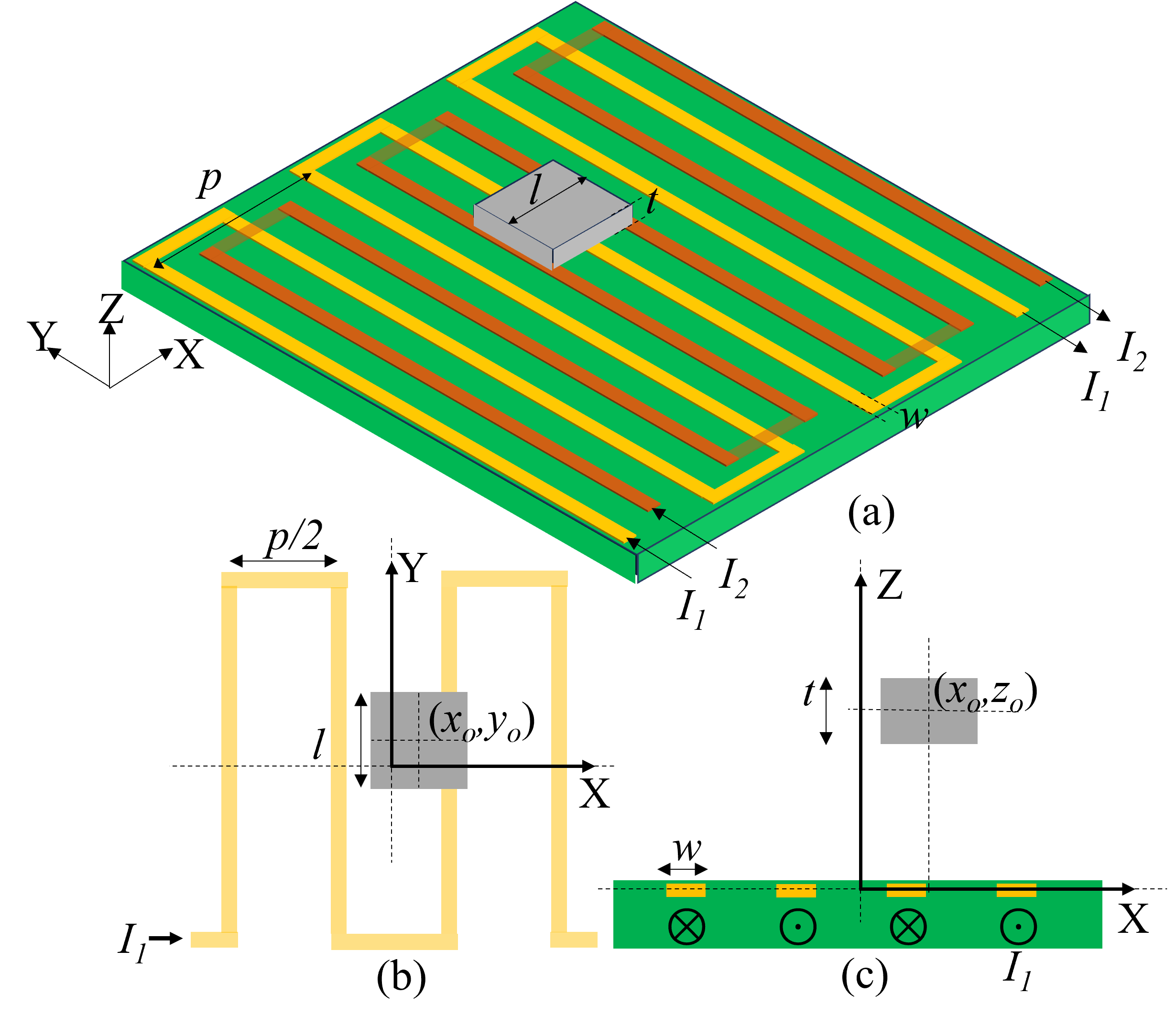}
    \par
    \end{centering}
    \caption{\small Schematics showing (a) Isometric view of the electromagnetic actuator, (b) Top-view and (c) Side view showing only one track and the magnet for modeling purposes. }
    \label{Schematic actuator}
\end{figure}

Figure \ref{Schematic actuator} (a) shows the schematic of the planar electromagnetic actuator. The actuator consists of two independent current-carrying tracks and a permanent magnet. Each track consists of multiple long straight segments parallel to the Y-axis, connected by short segments parallel to the X-axis. The resulting planar coil has a periodic structure with track width $w$ and pitch $p$. Since the thickness of the tracks will be significantly smaller than the other dimensions it has been neglected. The co-planar square-shaped permanent magnet has an edge length of $l$, a thickness of $t$ and is made of a material with uniform magnetisation $\textbf{M}= M\hat{z}\,$. 

For electromagnetic modeling, the magnet of volume $V$ is divided into multiple infinitesimal elements of volume $dV$ each with magnetic moment $\textbf{m}= MdV\hat{z}$. These elements experiences a force $\textbf{dF}$ due to the magnetic field  $\textbf{B} = $ [ $B_x$  $B_y$ $B_z$], generated by the current-carrying tracks, expressed as,

\begin{equation} \label{dF}
\textbf{dF} = (\textbf{m}.\nabla)\textbf{B} = \nabla(\textbf{m.B})
\end{equation}

The second form of Equation 1 is employed to calculate the electromagnetic force as $\nabla \times \textbf{B} = 0\,$ inside the magnetic volume. The total electromagnetic force is obtained by integrating the differential force over the volume of the magnet,

\begin{equation} \label{F}
\textbf{F}= \int_V M\nabla(B_z)dV.
\end{equation}

The schematic shown in Fig. \ref{Schematic actuator}(b-c) is employed for the evaluation of the electromagnetic force using Equation 2. First, the calculation of the magnetic field is discussed followed by the calculation of the electromagnetic force.

\subsubsection{Electromagnetic field}

The magnetic field due to track 1 is calculated away from the edges of the planar coil. To begin with, the parallel segments constituting track 1 are considered to be infinitely long and infinitesimally thin. The Z- magnetic field per unit current $b_{kz}$ due to the $k^{th}$ segment at a point $(x,z)$, obtained using Biot-Savart's law, is given by

\begin{equation} \label{bkz}
\begin{split}
b_{kz} & = (-1)^k\frac{\mu_o}{2\pi}\left[\frac{x+(1-2k)p/4}{(x+(1-2k)p/4)^2+z^2}\right], \\
& = \tilde b (\tilde x, \tilde z),
\end{split}
\end{equation}
where, $\tilde b (\tilde x, \tilde z) = (-1)^k \frac{\mu_o}{2\pi p}\left[\frac{\tilde x+(1-2k)/4}{(\tilde x+(1-2k)/4)^2 +(\tilde z)^2}\right]$, $\tilde{x}$ = $x/p$ and $\tilde{z}$ = $z/p$. The total magnetic field $b_{z1}$ due to all the segments constituting track 1 is obtained as 

\begin{equation} \label{bz1}
b_{z1} = \sum_{k=-N}^{N} \tilde b(\tilde x, \tilde z).
\end{equation}
 
If $N$ is sufficiently large, the magnetic field away from the edges of the PCB is a periodic function of $x$ with a fundamental period equal to the pitch $p$ of the track. Since $b_{z1}$ is an even function of $x$, it can be represented by the cosine terms $b_{z1} = b_0/2 + \sum_{n=1}^{\infty}b_ncos(2n\pi\tilde{x})$. Further, if $z \approx p $, then the periodic magnetic field per unit current can be represented by its first harmonic $b_{z1} \approx b_1cos(2\pi\tilde{x})$ as the contribution of higher harmonics is significantly small and hence can be neglected \cite{vikrant2022diamagnetically}. On substituting $\tilde{x}$ = 0 in Equation 4, the coefficient $b_1$   is determined to be

\begin{equation} \label{b1}
b_1 = \frac{\mu_0}{p}sech(2\pi\tilde{z}),
\end{equation}
where, $sech(2\pi\tilde{z})= 2/(e^{2\pi\tilde{z}}+ e^{-2\pi\tilde{z}})$.

It is to be noted that the magnetic field in Equation 4 is obtained assuming that the width of the segments constituting the track is negligible, i.e., $w \approx 0 $. However, the width of the segments can not be ignored if it is comparable to the track's pitch. To include the effect of finite width, each long segment of the track can be divided into $n$ infinitesimally thin segments of width $w/n\ll p $. The total magnetic field per unit current is then given by $B_{z1} = (b_1/n)\sum\limits_{k <n>}cos(\frac{2\pi}{p} (x-kw)) = (b_1/n)\frac{sin (\pi w/p)}{sin(\pi w/np)}cos(2\pi\tilde{x}) $. Further, in the limit when $n$ approaches infinity, we have 

\begin{equation} \label{Bz1}
B_{z1} = \frac{\mu_o}{p}sinc(\pi\tilde{w})sech(2\pi\tilde{z})cos(2\pi\tilde{x}),
\end{equation}
where $\tilde{w}$ = $w/p$. Equation 6 provides the magnetic field per unit current at any location $(x,z)$ away from the edges of the planar coil due to track 1. Since track 2 is identical to track 1 with an X- offset of $p/4$, the magnetic field per unit current due to track 2 is obtained by substituting $x$ with $x-p/4$ in equation 6

\begin{equation} \label{Bz2}
B_{z2} = \frac{\mu_o}{p}sinc(\pi\tilde{w})sech(2\pi\tilde{z})sin(2\pi\tilde{x}).
\end{equation}

The total Z-magnetic field due to the currents $I_1$ and $I_2$ through tracks 1 and 2 respectively is given by

\begin{equation} \label{Bz}
B_{z} = \frac{\mu_o I}{p}sinc(\pi\tilde{w})sech(2\pi\tilde{z})cos[2\pi(\tilde{x} - \theta)],
\end{equation}
where $I = \sqrt{I_1^2 +I_2^2} $ and $ \theta = (1/2\pi)\taninv(I_2/I_1)$.

\subsubsection{Electromagnetic force}
The electromagnetic force is obtained by calculating the volume integral in Equation 2. The schematic in Fig. \ref{Schematic actuator}(c) shows a magnet of uniform thickness $t$ with centre at $(x_o, z_o)$ and the elemental volume $dV = L(x-x_o)dxdz$. The electromagnetic X- and Z- forces acting on the magnet are given by,

\begin{equation} \label{Fx}
F_{x} = \int_{z_o-t/2}^{z_o+t/2}\int_{x_o-l/2}^{x_o+l/2} M\pdv{B_{z}}{x} \,L(x-x_o)dxdz,
\end{equation}

\begin{equation} \label{Fz}
F_{z} = \int_{z_o-t/2}^{z_o+t/2}\int_{x_o-l/2}^{x_o+l/2} M\pdv{B_{z}}{z} \,L(x-x_o)dxdz.
\end{equation}

Like electromagnetic fields, the forces will be calculated first for track 1. The obtained expression can then be used to calculate forces due to track 2 by substituting $x$ by $x-p/4$. For uniform magnetization M, the X-force per unit current due to track 1 is obtained using Equations 6 and 9

\begin{multline} \label{Fx1}
F_{x1} = -\frac{\mu_0 M}{p}sinc(\pi\tilde{w})\int_{x_o-l/2}^{x_o+l/2}\,L(x-x_o)sin(2\pi\tilde{x})dx\\
\int_{z_o-t/2}^{z_o+t/2}(2\pi/p)sech(2\pi\tilde{z})dz.
\end{multline}

The first integral in Equation 11 depends upon the in-plane orientation and the edge length of the magnet. For the orientation where the edges are parallel to the X- and Y- axes, $L(x-x_0)$ is equal to $l$. Thus, the integral of $lsin2\pi\tilde{x}$ over the given limits is $(\tilde{l}p^2/\pi)sin(\pi \tilde{l})sin(2\pi\tilde{x}_o),$ where  $\tilde{l} = l/p $ and $\tilde{x}_o= x_o/p $. The second integral depends upon the Z-position and the thickness of the magnet. The integral is evaluated to be $ gd[2\pi(\tilde{z}_o + \tilde{t}/2)]- gd[ 2\pi(\tilde{z}_o - \tilde{t}/2)],$ where $gd(\alpha)=tan^{-1}(sinh(\alpha))$, $\tilde{z}_o= z_o/p $ and $\tilde{t}= t/p$.   Finally, on substituting the variables $\tilde{x}_o$ and $\tilde{z}_o$ with $x$ and $z$ respectively, the force is found to be

\begin{equation} \label{Fx11}
F_{x1} = -\mu_o Mp\psi\phi sinc(\pi\tilde{w})sin(2\pi\tilde{x}),
\end{equation}
where $\psi = (\tilde{l}/\pi)sin(\pi \tilde{l})$ and $\phi = gd[2\pi(\tilde{z} + \tilde{t}/2)]- gd[ 2\pi(\tilde{z} - \tilde{t}/2)].$ The X-stiffness per unit current due to track 1 is obtained as

\begin{equation} \label{kx1}
\begin{split}
k_{x1} & =\pdv{F_{x1}}{x} \ , \\
& = -2\pi\mu_o M\psi \phi sinc(\pi\tilde{w})cos(2\pi\tilde{x}).
\end{split}
\end{equation}

Using the Taylor series expansion of $gd(2\pi \tilde{z})$, the expression $\phi$ can be approximated to $(2\pi \tilde{t})sech\left(2\pi \tilde{z}\right)$ provided the magnet is sufficiently thin,i.e., $t\ll p $. Thus for a thin magnet, the X- force and stiffness are given by 

\begin{equation} \label{Fx111}
F_{x1} = -\frac{2\pi\mu_o}{p^2}m_{eff}sinc(\pi\tilde{w})sech(2\pi\tilde{z})sin(2\pi\tilde{x}),
\end{equation}

\begin{equation} \label{kx11}
k_{x1} = -\frac{4\pi^2\mu_o} {p^3}m_{eff}sinc(\pi\tilde{w})sech(2\pi\tilde{z})cos(2\pi\tilde{x}),
\end{equation}

where, $m_{eff} = Mp^3\tilde{l}\tilde{t}sin(\pi \tilde{l})/\pi.$ Equation 14 shows that in a sinusoidal magnetic field, a finite dimension magnet experiences a force that is identical to the force experienced by a point dipole $m_{eff}$ positioned at the centre of the magnet. Similarly, for a thin magnet, the Z-force and the Z-stiffness per unit current due to track 1 are found to be 

\begin{equation} \label{Fz11}
\begin{split}
F_{z1} & = -\frac{2\pi\mu_o}{p^2}m_{eff}sinc(\pi\tilde{w})sech(2\pi\tilde{z})tanh(2\pi\tilde{z})\\
& \qquad    cos(2\pi\tilde{x}),
\end{split}
\end{equation}

\begin{equation} \label{kz1}
\begin{split}
k_{z1} & =\pdv{F_{z1}}{z} \ , \\
& = \frac{4\pi^2\mu_o}{p^3}m_{eff}sinc(\pi\tilde{w})sech(2\pi\tilde{z})(2sech^2(2\pi\tilde{z})-1)\\
& \qquad  cos(2\pi\tilde{x}).
\end{split}
\end{equation}

Equations 14-17 represent electromagnetic forces and stiffness per unit current due to track 1. Similarly, the forces and stiffness per unit current due to track 2 can be obtained by substituting $x$ with $x-p/4$. Therefore, the total electromagnetic force and stiffness due to the currents $I_1$ and $I_2$ through tracks 1 and 2 respectively are given by
   
\begin{equation} \label{Fxx}
F_x =-\frac{2\pi\mu_o I}{p^2}m_{eff}sinc(\pi\tilde{w})sech(2\pi\tilde{z}) sin[2\pi(\tilde{x}- \theta)],
\end{equation}

\begin{equation} \label{Fzz}
\begin{split}
F_z & = -\frac{2\pi\mu_o I}{p^2}m_{eff}sinc(\pi\tilde{w})sech(2\pi\tilde{z})tanh(2\pi\tilde{z})\\
& \qquad cos[2\pi(\tilde{x}-\theta)],
\end{split}
\end{equation}

\begin{equation} \label{kxx}
k_x = -\frac{4\pi^2\mu_o I}{p^3}m_{eff}sinc(\pi\tilde{w})sech(2\pi\tilde{z})cos[2\pi(\tilde{x}- \theta)],
\end{equation}

\begin{equation} \label{kzz}
\begin{split}
k_z & = \frac{4\pi^2\mu_o I}{p^3}m_{eff}sinc(\pi\tilde{w})sech(2\pi\tilde{z})(2sech^2(2\pi\tilde{z})-1)\\
& \qquad cos[2\pi(\tilde{x}-\theta)],
\end{split}
\end{equation}

where $I = \sqrt{I_1^2 +I_2^2} $ and $ \theta = (1/2\pi)\taninv(I_2/I_1)$.
Equations 18-21 provide the electromagnetic forces and stiffness for a thin square magnet whose centre is positioned at $(x,y,z)$. Using the currents $I_1 = I_ocos(\omega t)$ and $I_2 = I_osin(\omega t)$, an untethered magnet is stably trapped and actuated along the X-axis with a constant velocity $v = p\omega/2\pi$. Further, for the employed currents, the electromagnetic stiffness $k_x$ and $k_z$ of the trapped magnet during the X-motion will remain $-(4\sqrt{2}\pi^2\mu_o I_o/p^3) m_{eff}sinc(\pi\tilde{w})sech(2\pi\tilde{z})$ and $(4\sqrt{2}\pi^2\mu_o I_o/p^3) m_{eff}sinc(\pi\tilde{w})sech(2\pi\tilde{z})tanh(2\pi\tilde{z})$ respectively. 

It is to be noted that here we have only discussed trapping and actuation of the magnet along the X-axis by using a pair of tracks parallel to the Y-axis (Fig. \ref{Schematic actuator}). However, it is possible to electromagnetically trap and actuate the magnet along the Y-axis as well by using another pair of tracks which are parallel to the X-axis. A similar set of equations for forces and stiffness can be derived for motion along the Y-axis and is not being repeated here. Therefore, using the orthogonal pair of traces, the trapped magnet can be actuated over a large range in the XY-plane. Though the orthogonal pair of traces will stabilize the magnet in the XY-plane, it destabilizes the magnet along the Z-axis. This is reflected by the opposite signs of the electromagnetic stiffness $k_x$ and $k_z$ as shown in Equations 20 and 21 respectively. Therefore, the magnet has to be suspended either using flexures or by employing passive levitation techniques.  Next, the design parameters of the planar actuator required to maximize the electromagnetic force and stiffness will be discussed. 

\subsection{Optimization}
\begin{figure}[hbt!]
    \begin{centering}
        \includegraphics[width=0.4\paperwidth]{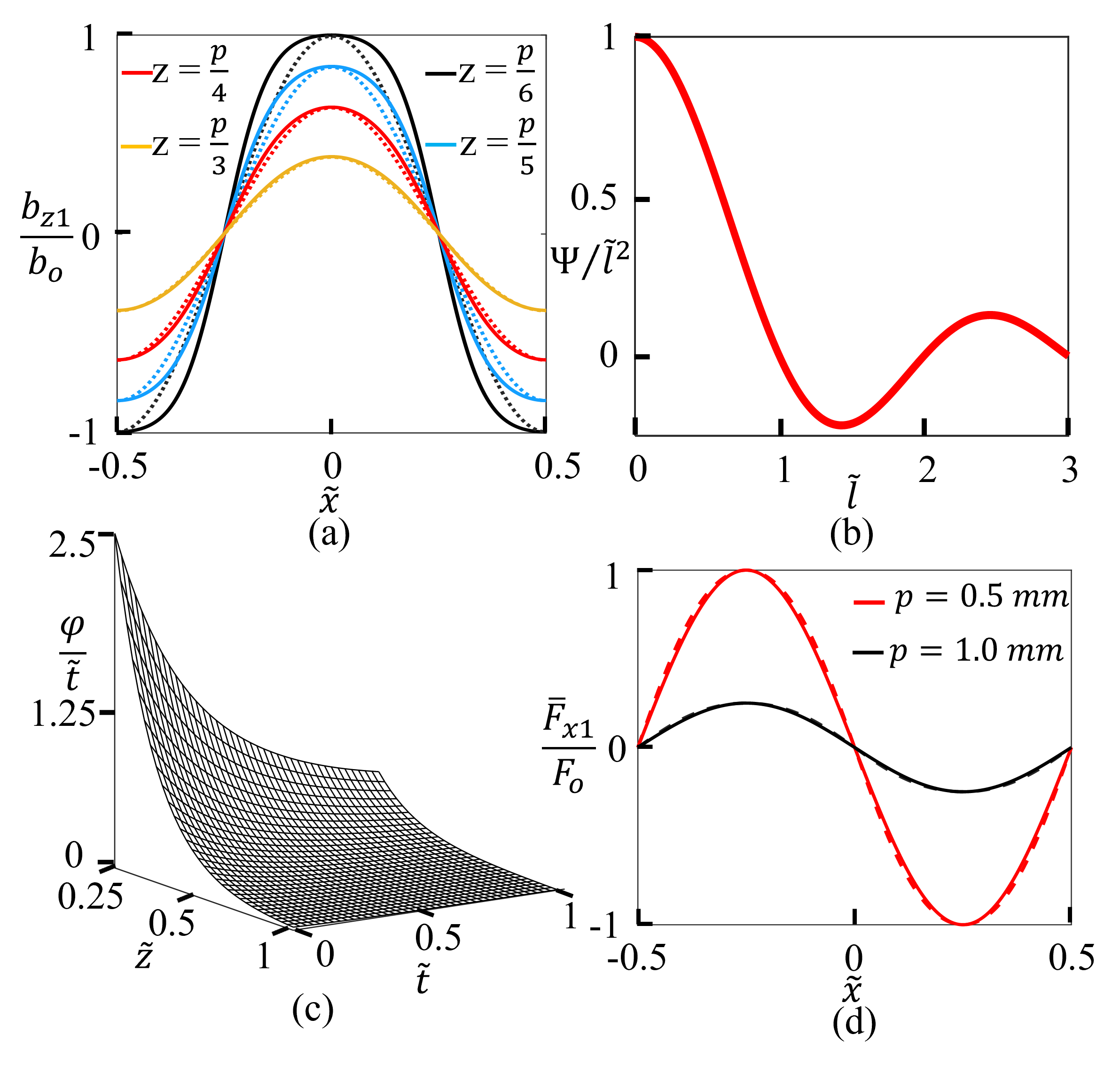}
    \par
    \end{centering}
    \caption{  \small Plots showing (a)  the electromagnetic field ${b}_{z1}$ and its first harmonic ${B}_{z1}$ represented by the solid and the dotted lines respectively, (b) $\frac{\psi}{\tilde{l^2}}$ as a function of $\tilde{l}$, (c)$\frac{\phi}{\tilde{t}}$ as a function of $\tilde{t}$ and $\tilde{z}$  and (d) the electromagnetic X-force $\overline{F}_{x1}$ represented by the solid line and the numerically obtained X-force shown in the dotted red line. The magnetic field and the force are normalized with their respective maximum value $b_o = $ 0.39 $mT/A$ and $F_o = $ 0.58 $MN/A.m^3$  The values of $N$, $w$ and $p$ are considered 100, 100 $\mu m$ and 1 $mm$ respectively for calculation of $b_{z1}$. The values of $\tilde{l}$, $\tilde{t}$ and $\tilde{z}$ are considered 0.5, 0.25 and 0.25 respectively for the calculation of X- force $\overline{F}_{x1}$. } 
    \label{Optimization}
\end{figure}

In this section, the design parameters of the planar coil and the untethered permanent magnet to achieve optimal performance will be determined. The goal is to select design parameters that provide a high electromagnetic force and high trapping stiffness per unit volume of the magnet. During the planar motion, the magnet remains suspended at a fixed Z-position above the PCB (Fig. \ref{Schematic actuator}(c)). To simplify the optimization process, it has been assumed that the minimum Z-position of the magnet $z_{min}\geq p/4$. The assumption is valid because the actuator employs a pyrolytic graphite plate or other substrate between the PCB and the magnet to achieve levitation or suspension \cite{hsu2017automated,vikrant2022diamagnetically,vikrant2023novel}. The thickness of this plate is usually on the order of one-quarter of a pitch, thus separating the magnet and the PCB by $p/4$. The advantage of considering $z \geq p/4$ is that the periodic magnetic field $b_{z1}$  can be represented by its first harmonic $B_{z1}$ as shown in Fig. \ref{Optimization}(a). The root mean square error between the periodic field and its first harmonic is 0.028 for $z = p/4 $.  Further, the error reduces as $z$ increases beyond $p/4$ and vice-versa. Due to the negligible error between the actual magnetic field and its first harmonic representation, the optimisation can be performed using the force and stiffness equations derived in the previous section.
 
 The design parameters required to maximize the in-plane force and stiffness per unit current can be obtained using Equations 12-13. Both the force per unit current and stiffness per unit current are normalised with the volume $l^2t$ of the magnet for comparison purposes

 \begin{equation} \label{Fx1f}
\overline F_{x1} = -\frac{\mu_o M}{p^2}\frac{\psi}{\tilde{l}^2}\frac{\phi}{\tilde{t}} sinc(\pi\tilde{w})sin(2\pi\tilde{x}),
\end{equation}

\begin{equation} \label{kx1f}
\overline k_{x1} = -\frac{2\pi\mu_o M}{p^3}\frac{\psi}{\tilde{l}^2}\frac{\phi}{\tilde{t}} sinc(\pi\tilde{w})cos(2\pi\tilde{x}),
\end{equation}

 where the factor $\psi/\tilde{l}^2 = sinc(\pi\tilde{l})$ depends only on $\tilde{l}$. Similarly, the factor  $\phi/\tilde{t}$ depends on $\tilde{z}_0$ and $\tilde{t}$. Fig. \ref{Optimization} (b) shows that the parameter $\psi/\tilde{l}^2$ is maximum when $l\ll p$. Similarly, Fig. \ref{Optimization} (c) shows that the parameter $\phi/\tilde{t}$ is maximum when $z = p/4$ and $t\ll p$. Therefore, a magnet with edge length and thickness much smaller than the pitch of the track will provide maximum normalized force and stiffness. Such a small magnet with an edge length of a few microns can be used for micro robotics applications because of its high natural frequency.

However, the small dimensions of the magnet prevent it from being used in other applications. Specifically, the small surface area of the magnet in the XY-plane limits the payload-carrying capacity of a levitation-based positioner
\cite{vikrant2022diamagnetically,hsu2018ferrofluid}. Similarly, the small volume limits the electromagnetic force and reduces the dynamic range of a flexure-guided beam integrated with the magnet.   One way to address these limitations is to increase the edge length $l$ of the magnet. Although increasing the surface area will improve the payload carrying capacity, it will reduce the bandwidth of a levitation-based positioner due to the reduction in stiffness per unit volume (Fig. \ref{Optimization} (b)). Similarly, the increase in edge length will increase the total electromagnetic force and hence the dynamic range but will reduce the bandwidth of the flexure guided magnetic assembly.  
 
\begin{figure}[hbt!]
    \begin{centering}
        \includegraphics[width=0.25\paperwidth]{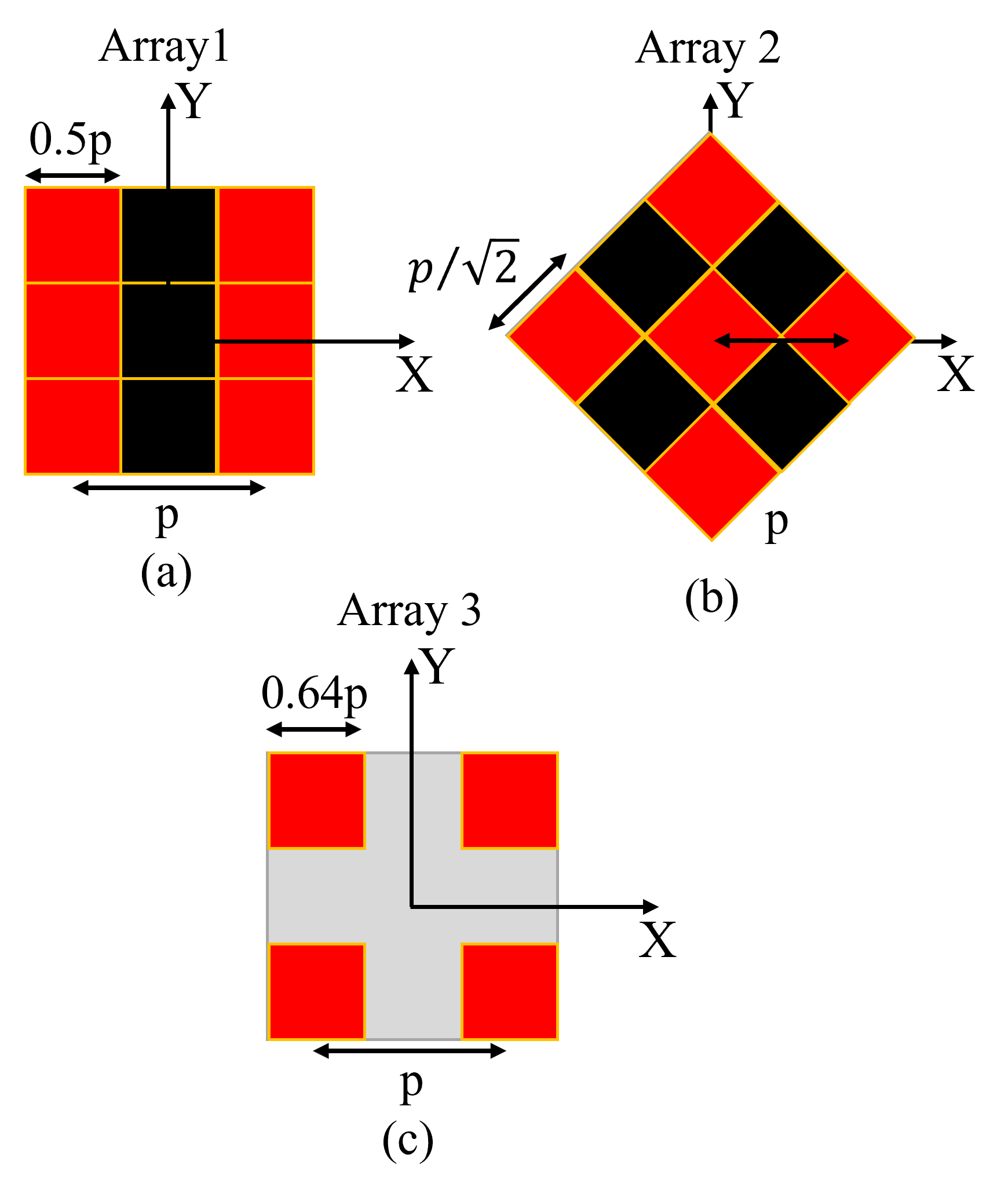}
    \par
    \end{centering}
    \caption{\small Schematics showing:  (a) Array 1 and (b) Array 2 designed entirely from permanent magnets, where the red and the black colour represents magnetic moment along positive and negative Z-axis respectively, (c) Array 3 designed using permanents magnets represented by red and non-magnetized material in between. The length of the magnets are considered as $p/2$, $p/\sqrt{2}$ and $0.64p$ for array 1, array 2 and array 3 respectively to provide maximum normalized force and stiffness.}
    \label{magnetic array}
\end{figure}

Instead of increasing the edge length of a single magnet, an alternative strategy exists to increase the surface area  without significantly decreasing the normalized electromagnetic force and stiffness.  The strategy is based on using an array of $N_m$ magnets, where magnets with identical magnetic moments are separated by a distance $p$ (Fig. \ref{magnetic array}). Since, the electromagnetic force and stiffness both are periodic with fundamental period $p$, each of magnets constituting the array will experience the same force and trapping stiffness. Thus, the array scales the area, the electromagnetic force and the electromagnetic stiffness all by the factor $N_m$ by leveraging the spatial periodicity of the magnetic field in the XY-plane.   

To scale the dimensions of the magnetic platform without compromising the normalized actuation force and trapping stiffness, three distinct array configurations can be designed (Fig. \ref{magnetic array}). The array configuration presented in Fig. \ref{magnetic array} (a) is tailored for generating X-motion without reducing the normalized X-force and X-stiffness. This is achieved by separating the magnets with similar magnetic moments by the distance $p$ only along the X-axis. Consequently all the magnets in the array 1 experiences the same X-force and X-stiffness, providing the maximum values of normalized force and stiffness (Table \ref{Table}).

While Array 1 provides the desired X-motion, it leads to non-optimal performance along Y-axis due to the non-periodic arrangement of the magnets across the column. One way to achieve optimal performance along both the axes is to use an orthogonal pairs of Array 1, where one array consists of magnets periodically arranged along X-axis and the other array consists of magnets periodically arranged along the Y-axis. Notably, previous applications have integrated such orthogonal pairs of arrays with a central platform to develop 2-axis micro and nano-positioners\cite{khan2011long,khan2017micro}. 

Instead of using orthogonal pairs of Array 1, either of the array configurations shown in Fig. \ref{magnetic array} (b-c) can be used to obtain optimal performance along both the axes.  In these arrays, magnets with identical magnetic moments are separated by $p$ along both the X- and Y-axis. This arrangement ensures that the electromagnetic force and stiffness scales by the same factor along both the axes with increase in the number of magnets. However, Array 2 and Array 3 provides $\approx 1.6$ and $\approx 3.4$ times lesser normalized stiffness compared to Array 1, respectively. Although Array 3 provides minimum normalized stiffness due to the use of non magnetic material between two consecutive magnets, it is the only configuration which can be micro fabricated as all the magnets have magnetic moments aligned in the same direction\cite{zhi2014planar,8808804}. 

It is worthwhile to note that although most of these array configurations have been employed previously, the dependence of force and stiffness upon design parameters have not been reported to the best of our knowledge. Here, we have obtained the closed form expressions for the normalized electromagnetic force and stiffness for all the three array configurations in terms of design parameters (Table \ref{Table}). Furthermore, optimization for all three configurations reveals that the normalized electromagnetic force and stiffness vary as  $\propto 1/p^2$ and $\propto 1/p^3$ respectively. Consequently, the performance can be significantly enhanced by reducing the pitch $p$ of the tracks.

\newcolumntype{M}[1]{>{\centering\arraybackslash}m{#1}}
\newcolumntype{N}{@{}m{0pt}@{}}

\begin{table}[ht]
\caption{ \small Comparison of the maximum values of normalized electromagnetic force and stiffness for 3 different array configurations. The width of the track and thickness of the magnets are considered small, i.e., $w \ll p$ and $t \ll p$.}
\begin{center}
\begin{tabular}{|M{2cm}|M{2cm}|M{2cm}|N}
\hline
 Array & max($\tilde{F}_{x1}$) & max($\tilde{k}_{x1}$)  &\\[20pt]
\hline
Array 1 & $\frac{5\mu_o M}{\pi p^2}$ & $\frac{10\mu_o M}{p^3}$ &\\[20pt]
\hline
Array 2 & $\frac{10\mu_o M}{\pi^2 p^2}$ & $\frac{20\mu_o M}{\pi p^3}$ &\\ [20pt]
\hline
Array 3 & $\frac{0.46\mu_o M}{p^2}$ & $\frac{2.9\mu_o M}{p^3}$ &\\ [20pt]
\hline

\end{tabular}
\end{center}
 \label{Table}
\end{table}

Finally, to show the accuracy of the optimization process, the forces obtained from derived closed-form expressions are compared with their numerically obtained values (Fig. \ref{Optimization}(d)). The plots shown in the solid lines are obtained using Equation 22 for two different pitches 0.5 mm and 1 mm. The plots shown in the dotted red lines are obtained by performing the integration in Equation 2 numerically for a single magnet whose edges are aligned parallel to the tracks. Since the root mean square error is 0.01, it shows that the optimization performed using the first harmonic representation for $ z \geq z_{min} $ gives accurate results. Furthermore, the plots show that the X-force increased by a factor of 4 as the pitch was reduced by a factor of 2. Therefore, the optimization shows that by designing a micro-scale magnetic array compared to the existing millimetre-scale magnetic array\cite{vikrant2022diamagnetically, hsu2017automated,vikrant2023novel}, the normalized stiffness can be increased by three orders of magnitude. This will lead to development of high bandwidth levitation based nano positioners as well high dynamic range and high bandwidth suspension based electromagnetic actuators.  

\section{Development and Evaluation}

In this section, first, the development of an experimental setup for the measurement of electromagnetic force will be discussed.  Next, the performance of the actuator for different design parameters will be compared. Finally, it will be shown that the performance can be improved significantly for the optimized design parameters obtained in the previous section.  

\begin{figure}[hbt!]
    \begin{centering}
        \includegraphics[width=0.40\paperwidth]{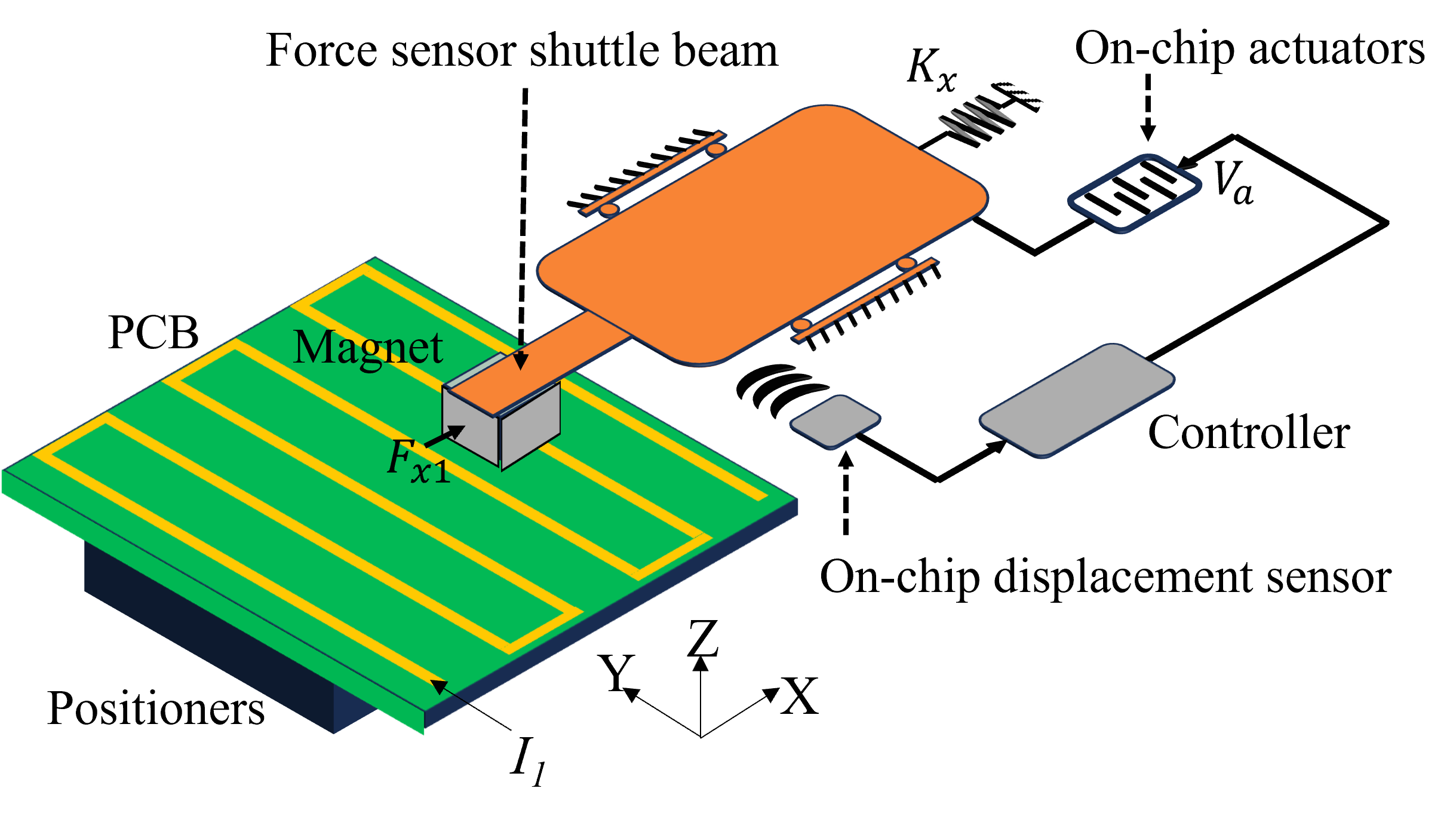}
    \par
    \end{centering}
    \caption{\small Schematic of the experimental set-up}
    \label{Experimental set-up}
\end{figure} 

The schematic of the experimental set-up used for the force measurement is shown in Fig. \ref{Experimental set-up}. The set-up comprises an electromagnetic actuator, a MEMS-based force sensor \cite{maroufi2018adjustable,DADKHAH2023103086} and positioners. The electromagnetic actuator consists of a single current-carrying track fabricated on a printed circuit board and a permanent magnet. It is to be noted that we have used a single track and a single magnet instead of two tracks and an array of magnets to keep the experiments simple. The current carrying track applies a force $F_{x1}$ on the magnet attached to the shuttle beam of the force sensor. The shuttle beam is integrated with a spring element with precisely known stiffness $K_x\gg k_x$ that deforms on the application of electromagnetic force $F_{x1}$. The deformation is measured using the on-chip displacement sensor. 

The applied force $F_{x1}$ can be estimated by multiplying the stiffness $K_x$  with the measured displacement of the shuttle beam. However, the measured force will be accurate only for small displacements of the shuttle beam due to the nonlinear response of the spring element for large deformation.  Further, the resolution of the force sensor will be affected by the external noise. To address both challenges, the force transducer is operated in a feedback control loop with a suitably designed controller $C(s)$.  First, the controller ensures that the spring element does not undergo a large deformation. This is achieved by applying an equal and opposite force $F_a = F_{x1}$ to the shuttle beam by utilizing on-chip actuators. The force applied by the on-chip actuators is given by $F_a = aV_a$ where $a$ is the actuation gain of the actuators and $V_a$ is the actuation voltage generated by the controller. Second, the controller also ensures that the external noise is rejected while tracking the applied force. Since the electromagnetic force $F_{x1}$ depends on the relative position of the magnet and the track, positioners (MTS25-Z8, Thorlabs) are employed to vary the position of the PCB while the magnet remains stationary.

\begin{figure}[hbt!]
    \begin{centering}
        \includegraphics[width=0.4\paperwidth]{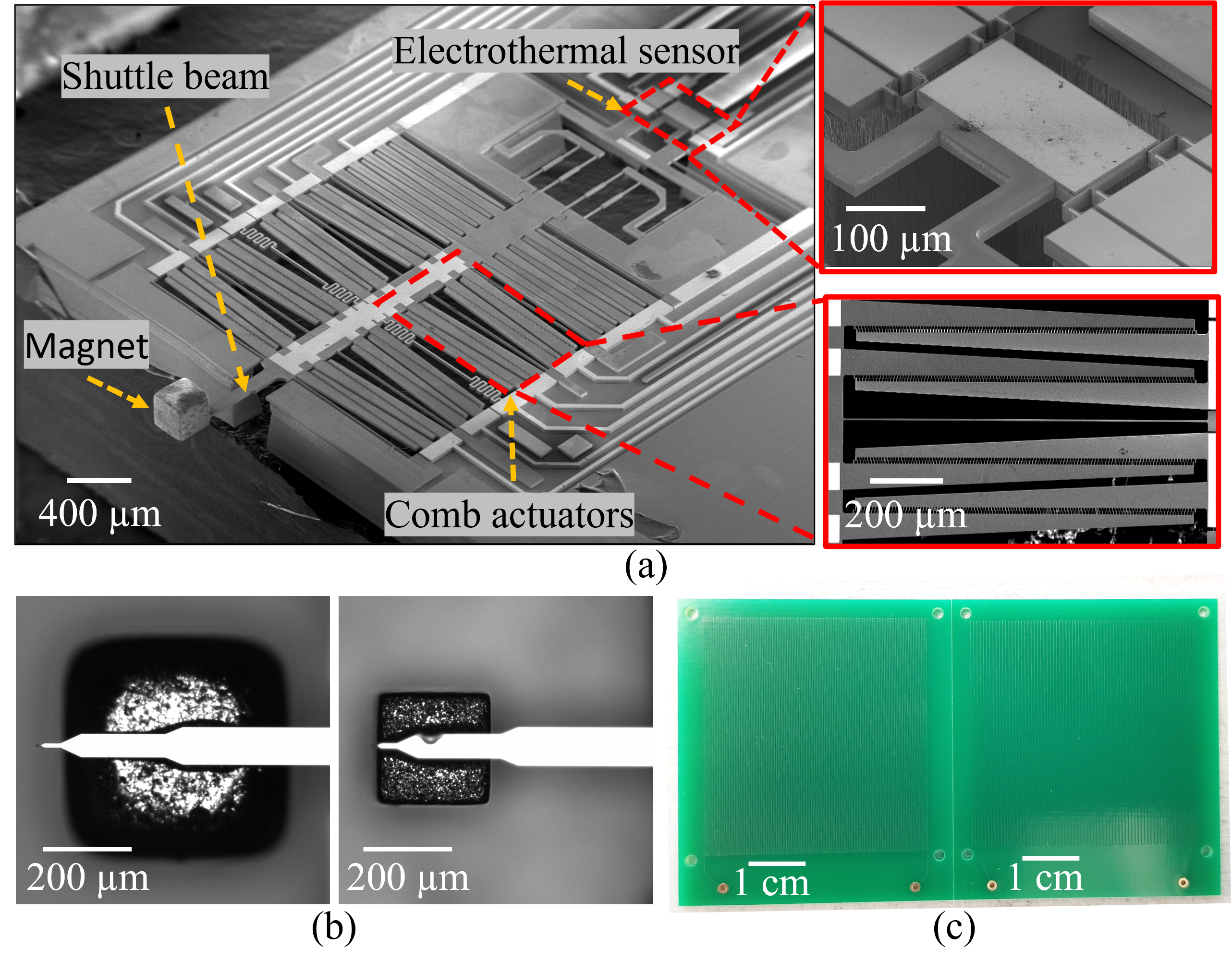}
    \par
    \end{centering}
    \caption{\small (a) Scanning Electron Microscope image of the force sensor with attached magnet, (b) Micro graphs showing two different magnets attached to the shuttle beams and (c) Image showing the two PCBs.}
    \label{Force sensor}
\end{figure}

 The force sensor and the electromagnetic actuator were developed and assembled in the configuration shown in Fig. \ref{Experimental set-up}. The force sensor comprising comb actuators and an electrothermal position sensor was fabricated using the MEMSCap's silicon-on-insulator microfabrication process (Fig. \ref{Force sensor} (a)).  The actuation gain and the measurement sensitivity of the comb actuator and electrothermal sensor were evaluated to be $ a = 9.2 \mu N/V $ and $ 0.30  V/\mu m $ respectively. The resolution of the electrothermal position sensor without the use of feedback is limited to $3.74 nm$ peak-to-peak primarily due to the presence of $60 Hz $ noise and its higher harmonics. Next, the stiffness of the flexures used for suspending the shuttle beam was experimentally evaluated to be $ K_x = 22.5 N/m $. Therefore, the open-loop measurement resolution of the force sensor is $84.15 nN$.
 
 Since the performance needs to be evaluated for different design parameters, three pairs of magnets and PCBs were selected for the experiments. While each of the three magnets has the same magnetization $ M = 10^6 A/m $, two of them are cubes and the third one is a thin cuboid magnet. Fig. \ref{Force sensor} (b) shows the cubic magnets with edge lengths $250 \mu m $ and $500 \mu m$. The thin cuboid magnet has an edge length of $500 \mu m $ and a thickness of $ 250 \mu m $. Three different PCBs with pitch $p$ of $2 mm$, $1 mm$ and $0.5 mm$ were fabricated for comparison purposes. The length, width and thickness of the long straight segments constituting the track for each of these PCBs were $50$ $mm$, $100$ $\mu m$ and $10$ $\mu m$ respectively. The image of the PCBs with pitch lengths $ 0.5 mm $ and $ 1 mm$  are shown in Fig. \ref{Force sensor} (c).
 
 \begin{figure}[hbt!]
    \begin{centering}
     \includegraphics[width=0.4\paperwidth]{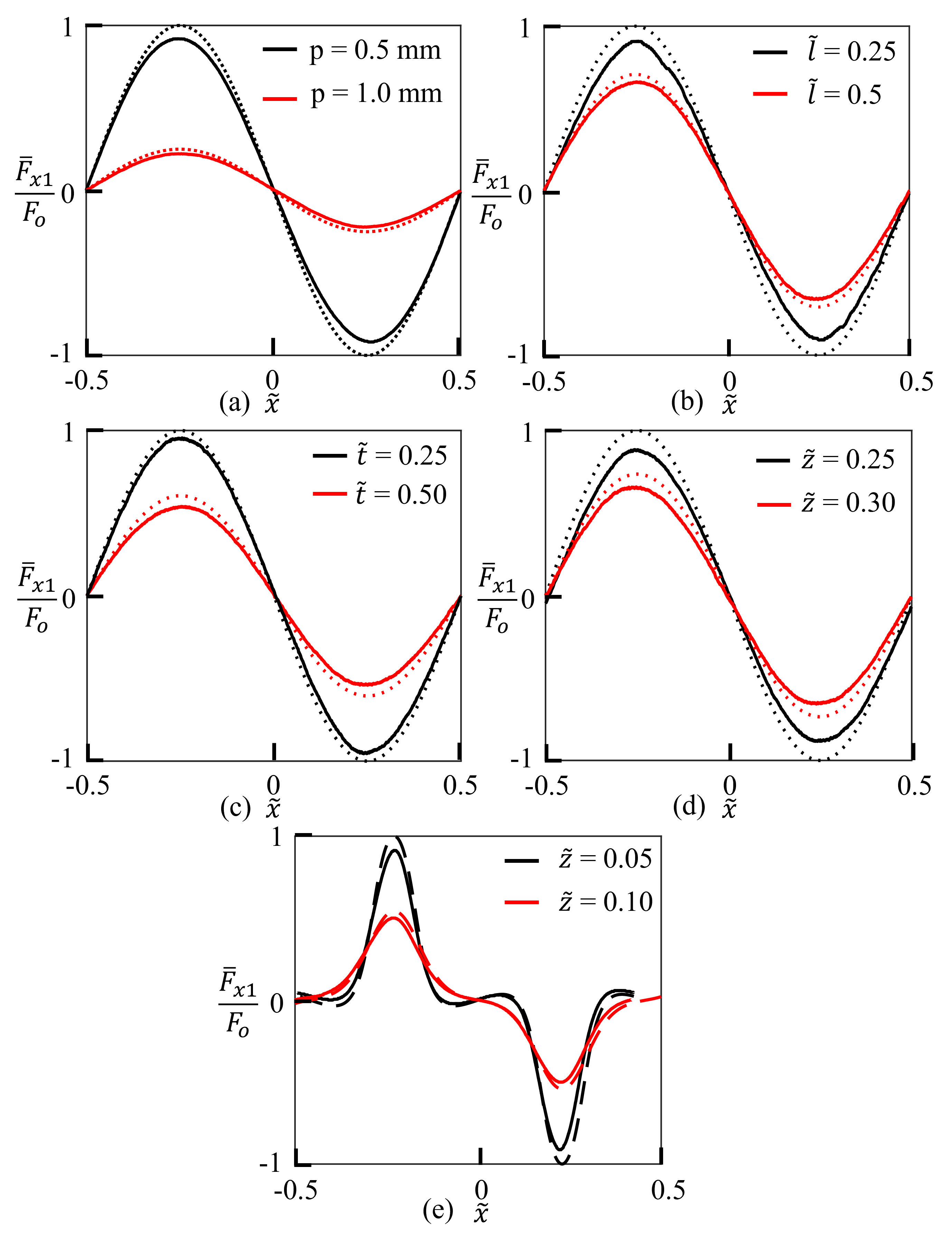}
    \par
    \end{centering}
    \caption{\small Plots showing the variation of electromagnetic forces with (a) pitch of the track while $\tilde{l}$, $\tilde{t}$ and $\tilde{z}$ were 0.5, 0.25 and 0.25 respectively, (b) length of the magnet while p, $\tilde{t}$ and $\tilde{z}$ were 1, 0.25 and 0.25 respectively, (c) thickness of the magnet while p, $\tilde{l}$ and $\tilde{z}$ were 1, 0.5 and 0.25 respectively, and (d-e) Z-position of the bottom surface magnet while p, $\tilde{l}$ and $\tilde{z}$ were 1, 0.5 and 0.25 respectively. The values of normalized peak force $F_o$ are $2.50 \frac{MN}{Am^3}$,$1.46 \frac{MN}{Am^3}$,$1.03 \frac{MN}{Am^3}$,$0.63 \frac{MN}{Am^3}$ and $3.81 \frac{MN}{Am^3}$ for the plots shown in (a), (b), (c), (d) and (e) respectively. Solid and dotted lines in each of the plots represent the experimental and analytical results respectively.}
    \label{Results}
 \end{figure}

 Finally, the experiments were performed wherein the energised PCBs were moved with a constant velocity of $50 \mu m/s$  along the X-axis and the force measurement was performed in closed loop. An integral controller $C(s) = 3.5/s$ was employed for tracking the slowly varying force while effectively rejecting the 60 Hz noise. With the designed controller, the resolution of the force sensor was found to be $4 nN$. Five different experiments were performed to show the dependence of the electromagnetic force $F_{x1}$ on design parameters $p$, $\tilde{l}$, $\tilde{t}$ and $\tilde{z}$.  In each of these experiments, one of the design parameters was varied while the other parameters were maintained constant. 
 
 In the first experiment, the pitch $p$ was reduced by a factor of two while $\tilde{l}$, $\tilde{t}$ and $\tilde{z}$ were maintained constant. The two-times reduction in pitch will increase the normalized force by a factor of 4 as shown in Fig. \ref{Results} (a). Similarly reducing the $\tilde{l}$ by a factor of two will increase $\psi/\tilde{l}^2$ and hence the force $F_{x1}$ by a factor 1.414 as shown in Fig. \ref{Results} (b). The next two experiments were performed to evaluate the effect of $\tilde{t}$ and $\tilde{z}$ (Fig. \ref{Results} (c-d)). It is worthwhile to note that the experimental and analytical results shown in the solid and dotted lines respectively matches  with a maximum error of less than 15\%. The analytical plots for all four experiments were obtained using Equation 22. 
 
 It is important to note that in all four experiments, the condition $z \geq p/4$ was ensured, which is often the case in the design of levitation-based micro robots and nanopositioners. However, there are specific applications where the magnet is directly placed on the PCB \cite{9143422}. In these scenarios, large in-plane forces are required to move the stage due to friction. To assess the variation of in-plane forces when $z \ll p$, a final experiment was conducted (Fig. \ref{Results} (e)). The plots illustrate that as $z$ decreases, the peak force increases, but the force becomes non-sinusoidal with reduced stiffness near the equilibrium position. These effects arise due to the  significant contribution of higher harmonics of the magnetic field as the Z-position reduces. The non-uniform reduced stiffness is likely to decrease the bandwidth of a levitation-based nanopositioner. Additionally, driving the levitating magnetic platform with uniform velocity will require complex current waveforms due to the nonlinear dependence of force on position.

 Therefore, to design a levitation-based electromagnetic actuator with high bandwidth, it is desirable to position the magnetic stage at $z\geq p/4$. Subsequently, the optimization study reported in the paper can be employed to design planar electromagnetic coils and magnetic stages where the pitch, edge length and thickness can be in micrometres. These actuators, that can be fabricated using existing microfabrication technologies, are expected to deliver force and stiffness per unit volume at least two to three orders of magnitude greater than current mesoscale electromagnetic actuators.
 
\section{Conclusion}

In this paper, we report optimized design parameters to improve the performance of a planar electromagnetic actuator based on the principle of linear motors. These actuators provide large-range planar motion with high positioning resolution when combined with passive levitation techniques. Here, we showed that the maximum load-generating capability and bandwidth of these actuators can be increased by several orders of magnitude using optimized design parameters. These design parameters were obtained from the derived closed-form expressions for the electromagnetic forces and stiffness. Finally, the electromagnetic actuator was developed and its load-applying capability and stiffness were evaluated. In particular, a MEMS-based force sensor which operated in a feedback loop to precisely measure the force was employed for the evaluation purpose. Notably, the obtained experimental results matched the analytical values with an error of less than 15\%. The obtained result paves the way for developing a micro-scale passive levitation-based magnetic array that can provide a high positioning bandwidth and will be part of our future work.




\bibliographystyle{ieeetr}
\bibliography{citations}

\begin{IEEEbiography}[{\includegraphics[width=1in,height=1.25in,clip,keepaspectratio]{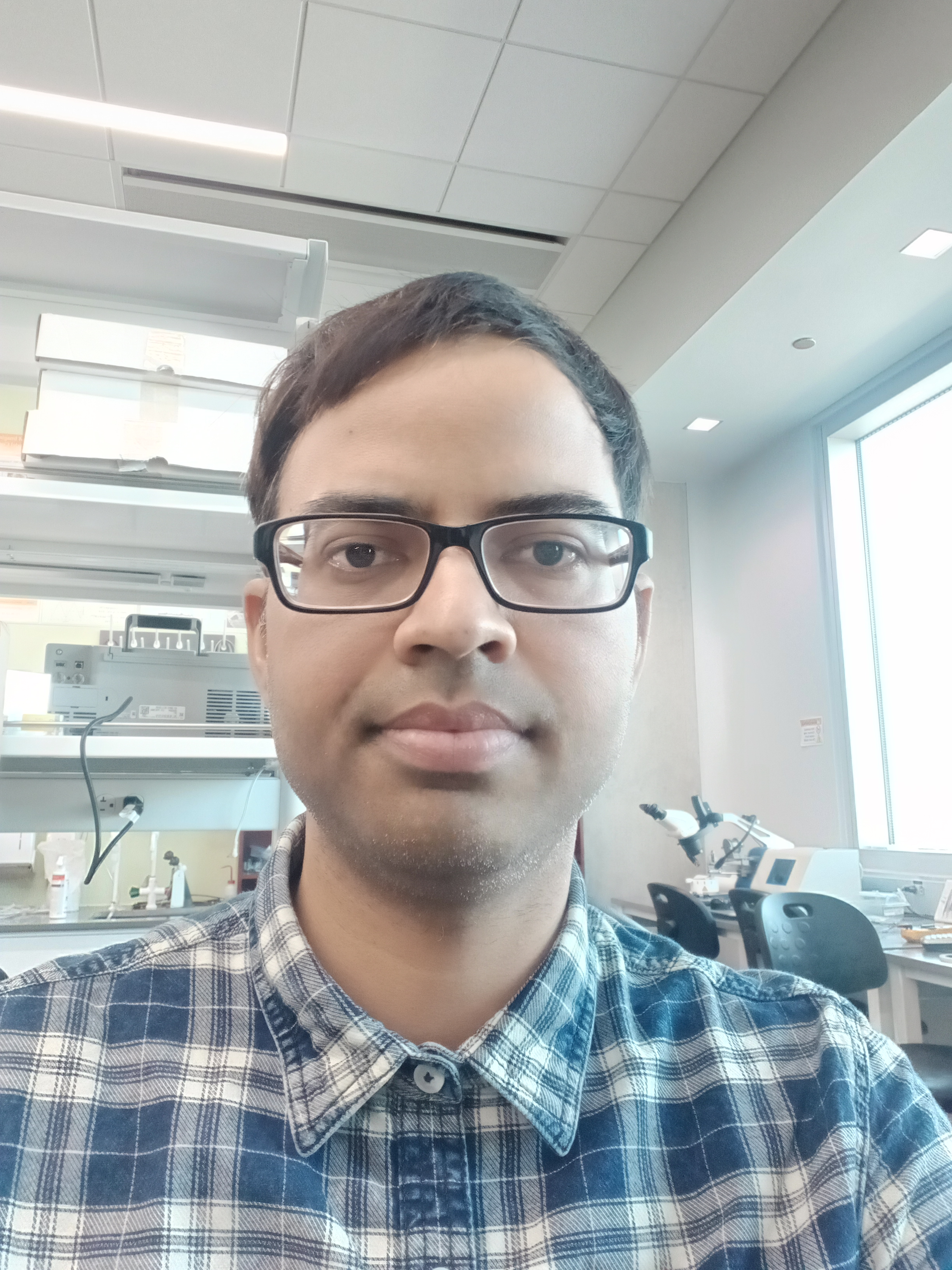}}]
{K.S. Vikrant} received his B.E. in Electronics and Instrumentation Engineering from the Institute of Technology and Management, Gwalior, in 2014 and both his MTech (Research) and PhD from the Indian Institute of Science, Bangalore in 2021. During his graduate research, he worked on precision actuation and measurement systems. He is currently a research scientist at the University of Texas at Dallas and is a member of the Laboratory for Dynamics and Control of Nanosystems. His current research interests include scanning probe microscopes, precision mechatronics, levitation-based nano-positioners, and micro-robotics.
\end{IEEEbiography}

\begin{IEEEbiography}[{\includegraphics[width=1in,height=1.25in,clip,keepaspectratio]{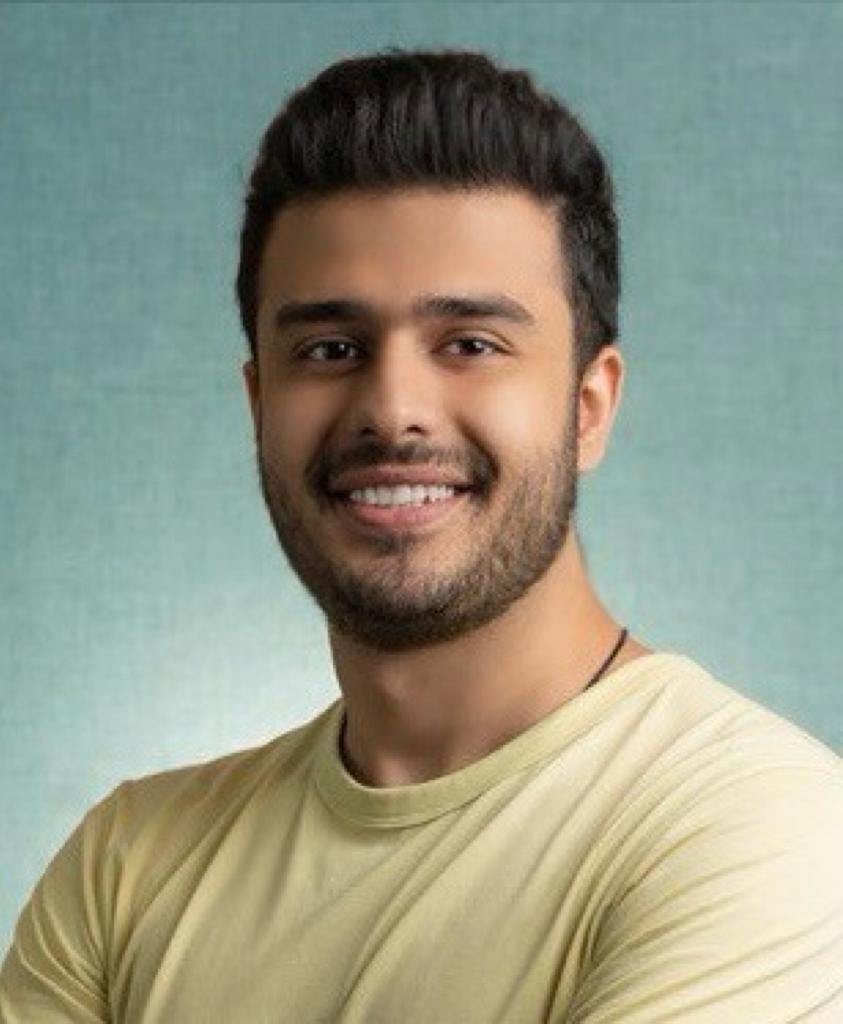}}]
{Diyako Dadkhah} received his B.Sc. in Mechanical Engineering from Sharif University of Technology, Tehran, Iran in 2021. He is currently pursuing the Ph.D. degree in mechanical engineering at the University of Texas at Dallas, and a member of the Laboratory for Dynamics and Control of Nanosystems (LDCN). His research interests are dynamic and control systems, MEMS, microfabrication, electronics, and signal processing.
\end{IEEEbiography}

\begin{IEEEbiography}[{\includegraphics[width=1in,height=1.25in,clip,keepaspectratio]{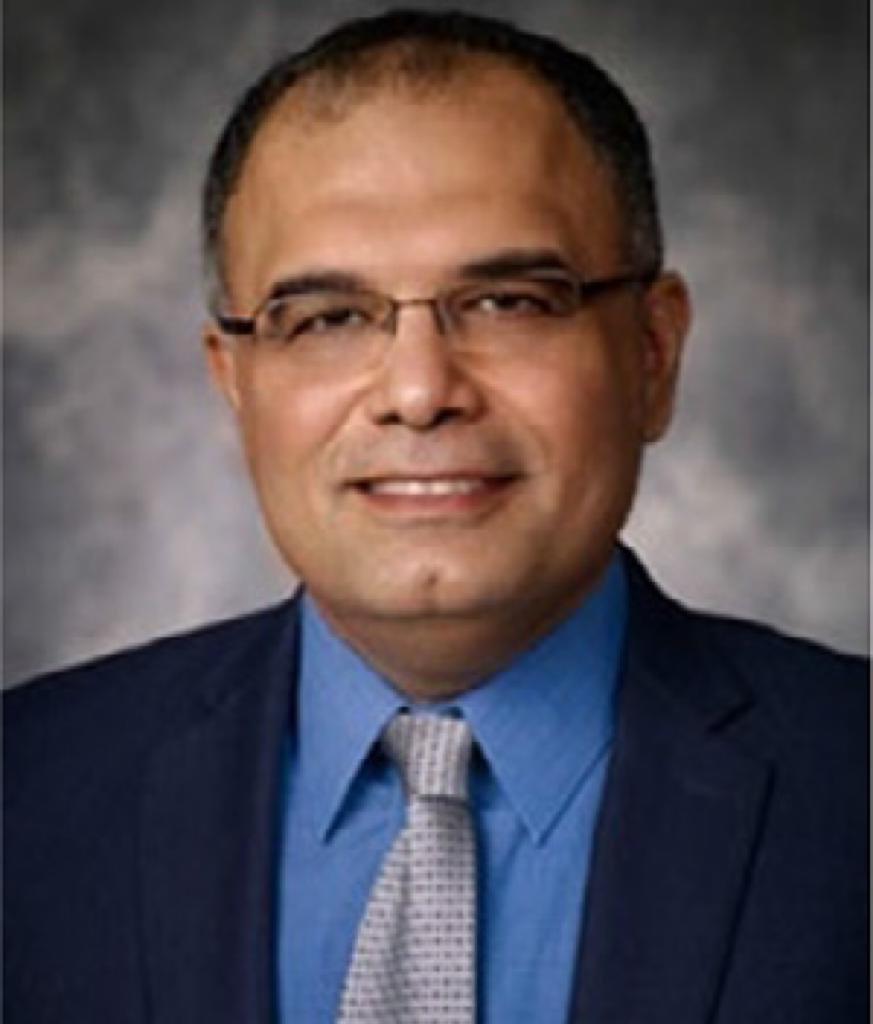}}]
{S. O. Reza Moheimani} is a professor and holds the James Von Ehr Distinguished Chair in Science and Technology in the Department of Systems Engineering at the University of Texas at Dallas with appointments in Electrical and Computer Engineering and Mechanical Engineering Departments. He is the founding Director of UTD Center for Atomically Precise Fabrication of Solid-State Quantum Devices and founder and Director of laboratory for Dynamics and Control of Nanosystems. He is a past Editor-in-Chief of Mechatronics (2016-2021), and a past associate editor of IEEE Transactions on Control Systems Technology, IEEE Transactions on Mechatronics and Control Engineering Practice. He received the Industrial achievement Award (IFAC, 2023), Nyquist Lecturer Award (ASME DSCD, 2022), Charles Stark Draper Innovative Practice Award (ASME DSCD, 2020), Nathaniel B. Nichols Medal (IFAC, 2014), IEEE Control Systems Technology Award (IEEE CSS, 2009) and IEEE Transactions on Control Systems Technology Outstanding Paper Award (IEEE CSS, 2007 and 2018). He is a Fellow of IEEE, IFAC, ASME, and Institute of Physics (UK).
Moheimani received the Ph.D. degree in Electrical Engineering from University of New South Wales, Australia in 1996. His current research interests include applications of control and estimation in high-precision mechatronic systems, high-speed scanning probe microscopy and atomically precise manufacturing. He is leading a multidisciplinary effort to develop new tools and methods for fabrication of solid-state quantum devices with atomic precision based on ultra-high vacuum scanning tunneling microscope.
\end{IEEEbiography}

\end{document}